\documentstyle[12pt,times]{article}

\begin{document}

\baselineskip 6mm
\renewcommand{\thefootnote}{\fnsymbol{footnote}}

\renewcommand{\baselinestretch}{1.24}	
\setlength{\jot}{6pt} 		
\renewcommand{\arraystretch}{1.24}   	

\def\IR{{\hbox{{\rm I}\kern-.2em\hbox{\rm R}}}}
\def\IZ{{\hbox{{\rm Z}\kern-.4em\hbox{\rm Z}}}}

\begin{titlepage}
\hfill\parbox{4cm}
{hep-th/0103264}

\vspace{15mm}
\centerline{\Large \bf Path integral formulation of Hodge 
duality on the brane}
\vspace{10mm}
\begin{center}
Sang-Ok Hahn\footnote{\tt hahn@newton.skku.ac.kr},  
Youngjai Kiem\footnote{\tt ykiem@newton.skku.ac.kr},
Yoonbai Kim\footnote{\tt yoonbai@skku.ac.kr},
and Phillial Oh\footnote{\tt ploh@newton.skku.ac.kr}\\[2mm] 
{\sl BK21 Physics Research Division and Institute of Basic
Science, Sungkyunkwan University, 
Suwon 440-746, Korea}
\end{center}
\thispagestyle{empty}
\vskip 40mm

\centerline{\bf ABSTRACT}
\vskip 5mm
\noindent
In the warped compactification with a single Randall-Sundrum brane,
a puzzling claim has been made that scalar fields can be bound 
to the brane but their Hodge dual higher-rank anti-symmetric tensors 
cannot.  By explicitly requiring the Hodge duality, a prescription
to resolve this puzzle was recently proposed by Duff and Liu.  In 
this note, we implement the Hodge duality via path integral 
formulation in the presence of the background gravity fields
of warped compactifications.  It is shown that the prescription
of Duff and Liu can be naturally understood within this framework.
\vspace{2cm}
\end{titlepage}

\baselineskip 7mm
\renewcommand{\thefootnote}{\arabic{footnote}}
\setcounter{footnote}{0}


The warped space-time geometry of Randall-Sundrum brane world 
in $(d+1)$-dimensions, whose
world-volume is $d$-dimensional, is described by the following
metric:
\begin{equation}
 ds^2 = \hat{g}_{MN} dX^M dX^N =
  \exp ( - 2 k |z| ) g_{\mu \nu} (x^{\mu} )  dx^{\mu} dx^{\nu} 
   + dz^2 ~. \label{metric}
\end{equation}
The coordinates of the bulk space-time
will be denoted as $X^M$, the coordinates
parallel to the brane, $x^{\mu}$,
and the direction perpendicular to the brane
will be parametrized by the coordinate $z$.  The brane is
assumed to be located at $z=0$. In the brane world scenario
with two branes \cite{RS1}, a natural geometric framework to deal with
the gauge hierarchy problem emerges as a consequence of the
the exponential warp factor in (\ref{metric}).   
After a wave-function renormalization, it 
produces an extremely tiny number $m_{\rm EW}/M_{\rm Planck} 
\sim 10^{-16}$ when the distance between the branes is
about 40 times the Planck scale.  The warp factor in 
the Jacobian $\sqrt{- \det \hat{g}_{MN} }$ allows 
normalizable gravitational zero modes localized on a single brane
when the range of $z$ is extended to $z=\infty$ \cite{RS2}.  

All these attractive features are due to the exponential suppression in the
large $z$ region coming from the covariant metric components $\hat{g}_{MN}$.
An opposite situation, however, might result when
the contravariant metric components $\hat{g}^{MN}$ are involved
in the classical action.  A particularly intriguing case 
involves the anti-symmetric tensor fields of 
various ranks and their Hodge duals.  Since the classical action integral 
over the $z$ direction is finite, a scalar field can be bound 
to the brane.  Higher rank anti-symmetric tensor fields that are Hodge dual
to scalars cannot however be bound to the brane, since the
similar integral diverges due to the large
number of contravariant metric components needed to contract
the field strengths \cite{KSS,Sil}.  To resolve this apparent 
violation of the Hodge duality, a proposal for a proper
Kaluza-Klein prescription was suggested by Duff and Liu \cite{duff}.
They found an appropriate Kaluza-Klein ansatz for higher
rank anti-symmetric tensor fields, which explicitly satisfies
the Hodge duality at the classical level.  
In the case of conventional (curved) space-time examples,
a fool-proof and straightforward approach to implement the Hodge duality 
is to formulate it in terms of path integrals \cite{KK}.  
In the same spirit, we implement the Hodge duality 
in the warped background geometry described by (\ref{metric}) 
via the path integral formalism.  We find that the prescription 
of \cite{duff} can be naturally obtained in this fashion.


The standard diffeomorphism invariant action for the
$p$-form anti-symmetric tensor gauge field 
$\hat{A} (X^M ) = \hat{A}_{[p]} (x^{\mu}, z)$ 
is\footnote{The material in this paragraph is based on 
our understanding of the discussions with Hyung Do Kim.}
\begin{equation}
 S_{F}  =  - \frac{1}{2 (p+1)!} \int d^{d+1}X \sqrt{- \hat{g}}
  \left( \hat{g}^{M_1 N_1} \cdots \hat{g}^{M_{p+1} N_{p+1}}
   \hat{F}_{M_1 \cdots M_{p+1}} \hat{F}_{N_1 \cdots N_{p+1}} 
  \right)  ~ .
\label{astart}
\end{equation}
To find a sensible Kaluza-Klein ansatz, we consider the
solutions to the parts of equations of motion
\begin{equation}
  \partial_z \left( \sqrt{-\hat{g}} \hat{g}^{zz}
 \hat{g}^{\mu_1 \nu_1} \cdots \hat{g}^{\mu_p \nu_p }
 \hat{F}_{z [ \nu_1 \cdots \nu_p ] } \right) = 0 ~ , 
\label{eom}
\end{equation}
which should be satisfied for the Kaluza-Klein zero modes.
Two consistent solutions to (\ref{eom})  
have been discussed in Refs.~\cite{KSS,Sil,duff}.  To be
specific, an ansatz is a zero-mode solution
\begin{equation}
  \hat{A}_{\mu_1 \cdots \mu_p}(x^{\mu}, z) =
A_{\mu_1 \cdots \mu_p} (x^{\mu}) \label{ansatz1} ~ ,
\end{equation}
and plugging it into (\ref{astart}) along with the background
metric produces the action
\begin{eqnarray}
   S_F & = & - \frac{1}{2 (p+1)!} \int dz \exp \left[ -2 k 
 \left(\frac{d}{2} - p -1 \right)  |z| \right] \nonumber \\
 & & \times  \int d^d x \sqrt{- g}
  \left( g^{\mu_1 \nu_1} \cdots g^{\mu_{p+1} \nu_{p+1}}
   F_{\mu_1 \cdots \mu_{p+1}} F_{\nu_1 \cdots \nu_{p+1}}
  \right) ~ . \label{action1}
\end{eqnarray}
The other ansatz is another consistent zero mode solution 
\begin{equation}
  \hat{A}_{z \mu_1 \cdots \mu_{p-1}}(x^{\mu}, z) =
  \exp \left[ -2 k \left( p -  \frac{d}{2} \right) k |z| \right] 
  A_{\mu_1 \cdots \mu_{p-1}} (x^{\mu}) ~ , \label{ansatz2}
\end{equation}
which leads to 
\begin{eqnarray}
   S_F & = & - \frac{1}{2 (p+1)!} \int dz \exp \left[ -2 k 
 \left(p - \frac{d}{2}  \right)  |z| \right] \nonumber \\
 & & \times  \int d^d x \sqrt{- g}
  \left( g^{\mu_1 \nu_1} \cdots g^{\mu_p \nu_p }
   F_{z \mu_1 \cdots \mu_p } F_{z \nu_1 \cdots \nu_p }
  \right) ~ , \label{action2}
\end{eqnarray}
upon following the same procedure.  
The factors containing the $z$-integral in (\ref{action1}) and
(\ref{action2}) are 
effectively the inverse coupling squared, $1/e^2_{\rm eff}$, in 
the $d$-dimensional brane theory.  The integral in 
(\ref{action1}) is convergent only for low values of $p$'s that satisfy
\begin{equation}
  p < \frac{d-2}{2} ~ .  \label{cond1}
\end{equation}
Therefore the bulk action (\ref{action1}) is sensible as the brane
action only for the $p$-form gauge fields satisfying the 
condition (\ref{cond1}).
Otherwise the effective coupling $e_{\rm eff}$ vanishes and the
gauge dynamics on the brane appears to be decoupled.  Under
the condition (\ref{cond1}), 
the $z$-integral in the action (\ref{action2})
diverges, making it problematic.   For the higher $p$-form gauge fields with 
\begin{equation}
   p  >  d/2 ~ , \label{cond2}
\end{equation}
the situation for each ansatz reverses itself.  While the $z$-integral
in the action (\ref{action1}) diverges, the $z$-integral in the other
action (\ref{action2}) converges.  
For the marginal cases which do not satisfy (\ref{cond1}) and (\ref{cond2}),
such as $p=(d-2)/2$ and $p = d/2$ for even $d$, the
divergence coming from the $z$-integral is linear in the cutoff
$z_0$, which translates to the logarithmic divergence 
$\log p_0$ in terms of 
the momentum cutoff $p_0$ of the $d$-dimensional brane theory.  
In \cite{Sil}, this was interpreted 
as the quantum charge screening effect in the brane theory (see 
however \cite{duff}).  In this paper, we will not consider these
marginal cases.    

This aspect appears rather puzzling if we use the ansatz (\ref{ansatz1})
for all values of $p$; the Hodge duality in the 
bulk space-time 
relates the bulk $p$-form gauge field with a $p^{\prime}$-form 
gauge field satisfying
\begin{equation}
  p^{\prime} = d - p - 1 ~ .
\end{equation}
When viewed from its Hodge dual side, the $p > d/2$
case should present no problem. Given 
this situation, the prescription of Duff and Liu \cite{duff}
is that one should use the ansatz (\ref{ansatz2}) for the 
$p$-form fields for $p > d/2$.
In what follows, we derive the action for the
$p$-form gauge field with $p > d/2$ from the action (\ref{action1}) 
of the $p$-form gauge field with $p < (d-2)/2$ by implementing
the Hodge duality in the path integral formalism~\cite{KK}.  
We find that this analysis precisely reproduces the prescription 
given by Duff and Liu \cite{duff}, which states that the ansatz
(\ref{ansatz1}) is relevant for $p < (d-2)/2 $ and the ansatz 
(\ref{ansatz2}) should be used for for $p > d/2$.  

The partition function for the brane world system includes
the following factor among others:
\begin{equation}
\int \left[ \prod_{X>0} {\cal D} \hat{A}_{[p]} (X) \right]
 \exp \left( i S_F \right) =  \int 
 \left[ \prod_{X>0} {\cal D} \hat{A}_{[p]} (X) \right]
 \exp \left( - \frac{i}{2} \int d^{d+1} X \hat{F}_I A_{IJ} 
 \hat{F}_J \right) ~ ,
\label{start}
\end{equation}
where 
\begin{equation}
 \frac{1}{2(p+1)!} \sqrt{- \hat{g}} \hat{g}^{\mu_1 \nu_1} 
  \cdots \hat{g}^{\mu_{p+1}
  \nu_{p+1}} \hat{F}_{\mu_1 \cdots \mu_{p+1}} 
 \hat{F}_{\nu_1 \cdots \nu_{p+1}} = \frac{1}{2} \hat{F}_I A_{IJ} 
  \hat{F}_J ~ .
\end{equation}
Each index $I$ collectively denotes the $(p+1)$-distinctive
directions chosen from the $d$ space-time directions parallel to
the brane: $I = \{ \mu_1^I , \cdots , \mu_{p+1}^I \}$. 
The matrix $A_{IJ}$ can thus
be written as
\begin{equation}
 A_{IJ} =  \sqrt{- \hat{g}} \hat{g}^{ \mu_1^I \mu_1^J} \cdots 
 \hat{g}^{ \mu_{p+1}^I \mu_{p+1}^J  } ~ .
\end{equation}
The set $\{ F_I | ~ {\rm all} ~ I \}$ has 
$D \equiv ~  _d C_{p+1}$ elements.  
The underlying bulk space-time geometry in the brane world
setup is invariant under the $\IZ_2$ transformation, whose action 
on $z$ is $z \rightarrow -z$, and the fixed points at $z=0$ 
correspond to the `world brane'.  Accordingly, the path 
integral in (\ref{start}) is over
the space-time points $X$ whose $z$-value is larger
than zero with an understanding that the fluctuations
of fields at $X(-z)$ are the `mirror images' of the fluctuations
at the point $X(z)$.  
For the gauge fields which do not have
the nonzero component along the $z$-direction, the zero mode
part of $\partial_z^2$ is a constant over the whole
range of the space $\IR^{1,d-1} \times \IR$, 
as shown in the standard Kaluza-Klein ansatz 
$\hat{A}_{\mu_1 \cdots \mu_p}(x^{\mu}, z) =
A_{\mu_1 \cdots \mu_p} (x^{\mu})$.  

The Hodge duality is implemented at the level of the path
integral by replacing the exponential function appearing in
(\ref{start}) via an identity
\begin{equation}
 \exp \left[  - \frac{i}{2} \int d^{d+1} X \hat{F}_I A_{IJ} 
 \hat{F}_J \right]
 =  \int \left[ \prod_{X >0 } \left[ \det ~ A_{IJ} (X) 
\right]^{-1}
 \prod_I {\cal D} \hat{\tilde{F}}_I (X)  \right] 
\label{gauge} \end{equation}
\[ \times 
 \exp \left[ \frac{i}{2} \int d^{d+1} X \hat{\tilde{F}}_I (A^{-1})_{IJ}
 \hat{\tilde{F}}_J + i \int d^{d+1} X \hat{\tilde{F}}_I 
 \hat{F}_I \right] ~ , \]
up to an overall numerical factor that we neglect from here on.
The symbol $\prod_{X>0}$ again signifies the fact that the
functional integral is over the bulk space-time points $X$ on
$\IR^{1,d-1} \times \IR^+ $, namely, 
the points $X$ whose $z$ coordinates satisfy $z>0$.  The
rationale behind this choice becomes clearer when we consider
the zero modes of the Hodge dual fields.  When the
Kaluza-Klein ansatz for the original gauge field is of the form 
in (\ref{ansatz1}), its non-vanishing 
Hodge dual field strength $\hat{\tilde{F}}$ 
necessarily involves the $z$-direction, or in other words,
the components $\hat{\tilde{F}}_{\mu_1 \cdots \mu_{p^{\prime}} z }
\ne 0$, where $p^{\prime} = d - p -1$.  Assuming their intrinsic
parity under $\IZ_2$ is even\footnote{There exist bulk higher
form gauge fields with negative intrinsic parity in
string/M theory.  For these fields, more refined arguments
are needed.}, the components  
$\hat{\tilde{F}}_{\mu_1 \cdots \mu_{p^{\prime}} z }$ transform
to $ - \hat{\tilde{F}}_{\mu_1 \cdots \mu_{p^{\prime}} z }$
under the $\IZ_2$ 
transformation $z \rightarrow -z$.  The components
$\hat{\tilde{F}}_{\mu_1 \cdots \mu_{p^{\prime}} z }$ thus 
contain the factor ${\rm sgn} (z)$; due to the discontinuity
at $z=0$, the components 
$\hat{\tilde{F}}_{\mu_1 \cdots \mu_{p^{\prime}} z }$  
are regular only in $\IR^{1,d-1} \times \IR^+ $.  Under our 
prescription, however, the path integral is only over the 
space-time region where the dual gauge fields are well-defined 
(regular), leading to (\ref{gauge}).  The consequence of 
this prescription is that the power of 
$\det  A_{IJ} (X)$ in (\ref{gauge}) is $-1$, unlike the 
case when the domain of the functional integration is over 
the whole simple covering space
$\IR^{1,d-1}  \times \IR$, which would lead to $-1/2$ in
(\ref{gauge}).   

The number of independent components of 
$\hat{\tilde{F}}_I$ is $_d C_{p^{\prime}}
= ~ _d C_{d - p -1} =  ~ _d C_{p+1} = D$, which is the same
number as the independent components of $\hat{F}_I$.  By a direct
combinatorics consideration, one can show that
\begin{equation}
 \frac{1}{2}\hat{\tilde{F}}_I (A^{-1})_{IJ}  \hat{\tilde{F}}_J = 
 \frac{1}{2(p^{\prime}+1)!} \sqrt{- \hat{g}} \hat{g}^{\mu_1 \nu_1} 
  \cdots \hat{g}^{\mu_{p^{\prime}}  
  \nu_{p^{\prime}} } \hat{g}^{zz} 
  \hat{\tilde{F}}_{\mu_1 \cdots \mu_{p^{\prime}} z} 
  \hat{\tilde{F}}_{\nu_1 \cdots \nu_{p^{\prime}} z}
\label{dualaa}
\end{equation}
and
\begin{equation}
 \det ~ A_{IJ} = ( - \hat{g} )^{ \frac{D}{2} - \frac{D}{d} (p+1 ) }
        = ( - \hat{g} )^{ \frac{D}{2d} (d - 2p -2 ) }
        = ( - \hat{g} )^{ \frac{D}{d} ( p^{\prime} - d/2 ) }
\end{equation}
using $\hat{g}_{zz} = 1$ and $\hat{g}_{\mu z} = 0$.
We observe that (\ref{dualaa}) is the standard kinetic
term for the $p^{\prime}$-form gauge field.
We use the classical approximation for the
path integral over the metric by inserting the background
metric (\ref{metric}) into the path integral (\ref{gauge}).
For the metric
of the type $~ \det ~ g = 1$, the measure part of (\ref{gauge}) 
can be rewritten as
\begin{equation}
 \left[ \prod_{X>0} \left[ \det ~ A_{IJ} (x) \right]^{-1}
 \prod_I {\cal D} \hat{\tilde{F}}_I (X)  \right] =
 \left[ \prod_{X>0} \prod_I  \exp \left[ 2 k (p^{\prime} - d/2 ) |z| 
  \right]
  {\cal D} \hat{\tilde{F}}_I (X) \right] \label{measure} ~ .
\end{equation}
The path integral over the ${\cal D} \hat{A}_{[p]}$ in
(\ref{start}), due to the $\hat{\tilde{F}}_I \hat{F}_I$ term in (\ref{gauge}),
imposes the usual constraint $ d \hat{\tilde{F}} = 0$ for the
functional integral, which
implies that only the field strength that can 
be written as $\hat{\tilde{F}} = d \hat{\tilde{A}}_{[p^{\prime}]}$ 
contributes to the path integral.  The only difference
between the conventional Hodge dual action in the flat
space and the case in consideration is the extra $z$-dependence
in the path integral measure (\ref{measure}). 
By writing, however,
\begin{equation}
 \hat{\tilde{A}}_{\mu_1 \cdots \mu_{p^{\prime}-1} z} = 
  \exp  \left[ - 2 k (p^{\prime} - d/2 ) |z| 
  \right] \tilde{A}_{\mu_1 \cdots \mu_{p^{\prime}-1} z } ~ ,
\label{redef}
\end{equation}
we can transform the measure part (\ref{measure}) into
\begin{equation}
 \left[ \prod_{X>0} \left[ \det ~ A_{IJ} (X) \right]^{-1}
 \prod_I {\cal D} \hat{\tilde{F}}_I (X)  \right] 
  = \left[ \prod_{X>0} \prod_I {\cal D} \tilde{F}_I (X)  \right] ~ .
\end{equation}
We note that $d \hat{\tilde{A}}_{[p^{\prime}]} = 0$
automatically implies $d \tilde{A}_{[p^{\prime}]} = 0$.
Our main result is that (\ref{redef}) is 
precisely the Kaluza-Klein ansatz appearing in (7) 
of Ref.~\cite{duff} for the $p$-form gauge fields with $p > d/2$. 


In this note, we have shown that the higher form gauge fields
that are Hodge dual to the gauge fields that can survive on
the brane also survive on the brane.  Restricting our attention
only to `fundamental' gauge fields that have the conventional
Kaluza-Klein ansatz, however, the condition (\ref{cond1})
still applies.  It is amusing to note that the `fundamental'
higher form gauge fields do not live on the brane world,
which happens to be the property of our world.

Our analysis can be generalized in various directions.
One can, for example, repeat the same type of analysis in the
context of higher (transversal) dimensional warped compactifications,
where the background metric is given by
\begin{equation}\label{metric2}
 ds^2 = \hat{g}_{MN} dX^M dX^N =
  \exp ( - 2 k r ) g_{\mu \nu} (x^{\mu} )  dx^{\mu} dx^{\nu} 
   + dr^{2}+C^{2}(r)d\Omega_{N-1} ~ .
\end{equation}
These background geometries are typically classified into 
two categories; the cigar geometries of constant 
radius, $C(r)=C_{\infty}$, and bottle-neck geometries of exponentially decreasing 
radius, $C(r)=e^{-kr}$~\cite{Gre}.  These two cases produce different 
classical action and different Jacobian factor in the path integral 
measure after taking Hodge duality.  Comparing them and finding
correct Kaluza-Klein ansatz should be an interesting issue \cite{HL}.
Secondly, in view of the string-inspired model constructions,
it is important to consider the bulk gauge fields with odd intrinsic 
parity.   

\section*{Acknowledgements}

The authors would like to thank Jin-Ho Cho and Hyung Do Kim 
for useful discussion.  We are also grateful to M.J. Duff and
J.T. Liu for  encouraging correspondence.  This work is supported by 
No. 2000-1-11200-001-3 from the Basic Research
Program of the Korea Science $\&$ Engineering Foundation
and Korea Research Center for Theoretical Physics and 
Chemistry (Y. Kim).

\end{document}